\documentclass[aip,jcp,reprint,noshowkeys,superscriptaddress]{revtex4-1}
\usepackage{graphicx,dcolumn,xcolor,microtype,multirow,amscd,amsmath,amssymb,amsfonts,physics,wrapfig,xspace}
\usepackage[version=4]{mhchem}
\usepackage{chemfig}
\usepackage{siunitx}
\DeclareSIUnit[number-unit-product = {\,}]
\cal{cal}
\DeclareSIUnit\kcal{\kilo\cal}

\DeclareSIUnit\debye{D}
\usepackage{hyperref}
\hypersetup{
    colorlinks,
    linkcolor={red!50!black},
    citecolor={red!70!black},
    urlcolor={red!80!black}
}

\newcommand{\alert}[1]{\textcolor{black}{#1}}
\usepackage[normalem]{ulem}

\newcommand{\SupMat}{\textcolor{blue}{supporting information}\xspace}

\newcommand{\fnm}{\footnotemark}
\newcommand{\fnt}{\footnotetext}
\newcommand{\tabc}[1]{\multicolumn{1}{c}{#1}}

\newcommand{\Dhf}{HF-$\Delta$CCSD\xspace}
\newcommand{\Doo}{oo-$\Delta$CCSD\xspace}
\newcommand{\eom}{EOM-CCSD\xspace}

\newcommand{\Phic}{\Phi_{\text{core}}}

\newcommand{\hH}{\Hat{H}}
\newcommand{\hHeff}{\Hat{H}^\text{eff}}
\newcommand{\hP}{\Hat{P}}
\newcommand{\hOmega}{\Hat{\Omega}}

\newcommand{\hh}{\Hat{h}}
\newcommand{\hf}{\Hat{f}}
\newcommand{\bH}{\Bar{H}}
\newcommand{\hT}{\Hat{T}}
\newcommand{\hR}{\Hat{R}}

\newcommand{\cre}[1]{\Hat{a}_{#1}^{\dag}}
\newcommand{\ani}[1]{\Hat{a}_{#1}^{}}

\newcommand{\hi}[1]{\Hat{#1}^{}}
\newcommand{\hid}[1]{\Hat{#1}^{\dag}}

\newcommand{\ES}{E^{\text{S}}}
\newcommand{\ET}{E^{\text{T}}}

\newcommand{\Heff}{H^\text{eff}}

\newcommand{\mM}{\mathcal{M}}

\newcommand{\LCPQ}{Laboratoire de Chimie et Physique Quantiques (UMR 5626), Universit\'e de Toulouse, CNRS, UPS, France}
\newcommand{\CEISAM}{Nantes Universit\'e, CNRS,  CEISAM UMR 6230, F-44000 Nantes, France}
\newcommand{\IUF}{Institut Universitaire de France (IUF), F-75005 Paris, France}

\begin{document}

\title{State-Specific Coupled-Cluster Methods for Excited States}

\author{Yann \surname{Damour}}
	\email{yann.damour@irsamc.ups-tlse.fr}
	\affiliation{\LCPQ}
\author{Anthony \surname{Scemama}}
%	\email{scemama@irsamc.ups-tlse.fr}
	\affiliation{\LCPQ}
\author{Denis \surname{Jacquemin}}
%	\email{denis.jacquemin@univ-nantes.fr}
	\affiliation{\CEISAM}
	\affiliation{\IUF}
\author{F\'abris \surname{Kossoski}}
	\email{fabris.kossoski@irsamc.ups-tlse.fr}
	\affiliation{\LCPQ}
\author{Pierre-Fran\c{c}ois \surname{Loos}}
	\email{loos@irsamc.ups-tlse.fr}
	\affiliation{\LCPQ}

\begin{abstract}
We reexamine $\Delta$CCSD, a state-specific coupled-cluster (CC)
with single and double excitations (CCSD) approach that targets
excited states through the utilization of non-Aufbau determinants. This methodology
is particularly efficient when dealing with doubly excited states, a domain where
the standard equation-of-motion CCSD (EOM-CCSD) formalism falls
short. Our goal here is to evaluate the effectiveness of $\Delta$CCSD
when applied to other types of excited states, comparing its
consistency and accuracy with EOM-CCSD. To this end,
we report a benchmark on excitation energies computed with the $\Delta$CCSD and EOM-CCSD methods,
for a set of molecular excited-state energies that encompasses not only doubly excited states but
also doublet-doublet transitions and (singlet and triplet) singly-excited states of closed-shell systems.
In the latter case, we rely on a minimalist version of multireference CC known as
the two-determinant CCSD method to compute the excited states. Our dataset,
consisting of 276 excited states stemming from the \textsc{quest} database
[\href{https://doi.org/10.1002/wcms.1517}{V\'eril \textit{et al.},
\textit{WIREs Comput. Mol. Sci.} \textbf{2021}, 11, e1517}], provides a significant
base to draw general conclusions concerning the accuracy of $\Delta$CCSD.
Except for the doubly-excited states, we found that $\Delta$CCSD underperforms EOM-CCSD.
For doublet-doublet transitions, the difference between the mean absolute errors (MAEs) of the two methodologies (of \SI{0.10}{\eV} and \SI{0.07}{\eV})
is less pronounced than that obtained for singly-excited states of closed-shell systems (MAEs of \SI{0.15}{\eV} and \SI{0.08}{\eV}).
This discrepancy is largely attributed to a greater number of excited states in the latter set exhibiting multiconfigurational characters,
which are more challenging for $\Delta$CCSD.
We also found typically small improvements by employing state-specific optimized orbitals.
\bigskip
\begin{center}
	\boxed{\includegraphics[width=0.5\linewidth]{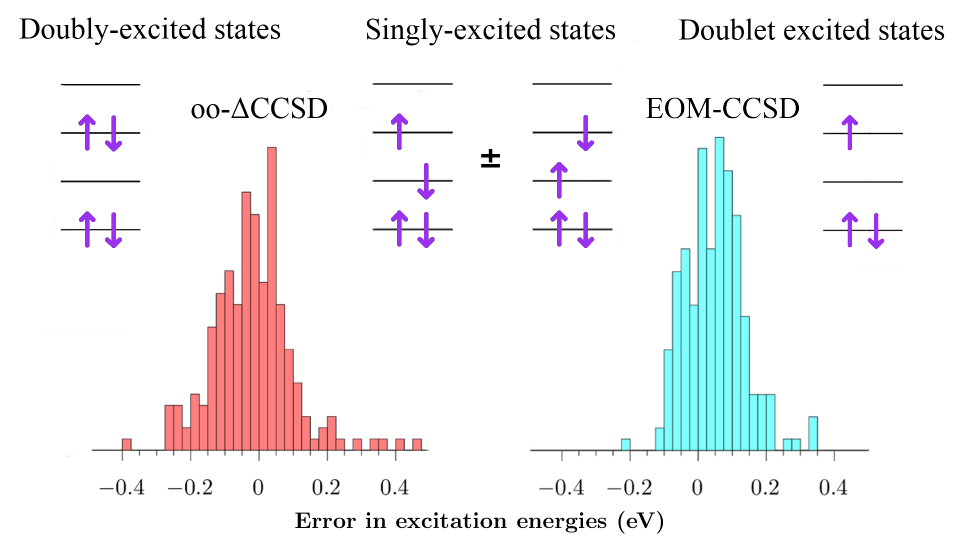}}
\end{center}
\bigskip
\end{abstract}

\maketitle

%%%%%%%%%%%%%%%%%%%%%%%%%%%%%%%%%%%%%%%%%%%%%%%
\section{Introduction}
\label{sec:intro}
%%%%%%%%%%%%%%%%%%%%%%%%%%%%%%%%%%%%%%%%%%%%%%%

Over the past few years, significant advances have been made towards
the accurate description of molecular excited states, with a particular
emphasis on vertical excitation energies from the electronic ground state.
These advances have been realized by \alert{solving the Schr\"odinger equation in
a basis of many-electron functions} using a set of judicious and appropriate
approximations. In such a way, numerous innovative methods have emerged,
each characterized by distinct strategies designed to describe the ground
and excited states. \cite{Roos_1996a,Piecuch_2002,Dreuw_2005,Krylov_2006,Sneskov_2012,Gonzales_2012,Laurent_2013,Adamo_2013,Ghosh_2018,Blase_2020,Loos_2020d}

The prominent density-functional theory (DFT) \cite{Hohenberg_1964,Kohn_1965,Parr_1989,Teale_2022}
provides a framework to study electronic excited states through time-dependent
DFT, \cite{Runge_1984,Burke_2005,Casida_2012,Huix-Rotllant_2020} by applying
the linear-response formalism on top of a ground-state calculation. Similarly,
single-reference coupled-cluster (SRCC)\cite{Cizek_1966,Cizek_1969,Paldus_1992,Crawford_2000,Bartlett_2007,Shavitt_2009}
methods compute excitation energies based on the ground-state amplitudes.
This is done by either constructing and diagonalizing the similarity-transformed
Hamiltonian in equation-of-motion coupled-cluster (EOM-CC) \cite{Rowe_1968a,Emrich_1981,Sekino_1984,Geertsen_1989,Stanton_1993a,Comeau_1993,Watts_1994}
or by examining the poles of the linear-response function in linear-response
coupled-cluster (LRCC).\cite{Monkhorst_1977,Dalgaard_1983,Sekino_1984,Koch_1990c,Koch_1990a}
Consequently, the excited states provided by these frameworks are inherently
biased toward the ground state upon which they are built. 
As a direct consequence, \alert{even if EOM- and LR-CC are in principle exact,
the accuracy of such approaches is known to be quite poor for
certain classes of excited states where the truncation of the excitation
operator is too severe compared to the degree of the excitation.
This is typically the case for doubly-excited states when EOM/LR-CC is
restricted to single and double excitations.} \cite{Watson_2012,Shu_2017,Barca_2018b,Loos_2019,Ravi_2022,doCasal_2023}

Nevertheless, other strategies can be used with different methods to
address this limitation. In configuration interaction (CI), \cite{Szabo_book}
excited-state energies are obtained through the diagonalization of the
Hamiltonian matrix within a basis of Slater determinants. However, since
the results are highly dependent on the underlying orbitals, their choice
is critical. To deal with this issue,
the states can be treated in a state-averaged approach, i.e., using the same
set of orbitals for the states of interest, thereby treating them on an equal
footing. \cite{Werner_1981,Diffenderfer_1982} Excitation energies can then be straightforwardly extracted as energy
differences between these states. Alternatively, one can employ a state-specific
approach, which requires to use of orbitals optimally designed for each of
the states of interest and implies separate calculations for each of them.
When seeking to evaluate the excitation energy between two states, it is
intuitive to expect the state-specific approach to yield more accurate
results. Nonetheless, certain issues might be encountered, such as keeping
track of the targeted state because of possible root-flipping.\cite{Docken_1972,Golab_1983}
Similarly, variational collapse may happen while trying to follow the
potential energy surface of a given excited state. Thus, even if
such problems can occur, by adopting one of these two approaches (state-averaged
or state-specific), the calculations are no longer subject to an orbital-related
bias toward the ground state.

At the self-consistent-field (SCF) level, such as Hartree-Fock (HF) or DFT,
the well-known $\Delta$SCF method employs non-Aufbau determinants with orbitals
that are optimized in a state-specific manner to describe excited states.\cite{Ziegler_1977,Gilbert_2008,Kowalczyk_2011}
A considerable body of research has been dedicated to the $\Delta$SCF family
of methods, yielding promising results, notably for doubly-excited and
charge-transfer states.\cite{Filatov_1999,Kowalczyk_2013,Levi_2020a,Levi_2020b,Hait_2020,Hait_2021,Cunha_2022,Toffoli_2022,Shea_2018,Barca_2018,Shea_2020,Hardikar_2020,Carter-Fenk_2020,Wibowo_2023,Schmerwitz_2022,Schmerwitz_2023}
Multireference self-consistent-field \cite{Das_1966,Das_1973,Roos_1980a,Roos_1980b,Malmqvist_1990,Roos_2016}
and multireference perturbation theory \cite{Andersson_1990,Andersson_1992,Finley_1998,Angeli_2001,Angeli_2002,Battaglia_2023}
methods have also been extensively applied in a state-averaged or state-specific way. \cite{Shea_2018,Tran_2019,Tran_2020,Burton_2021,Burton_2022,Marie_2023}
In particular, some of us have recently employed a state-specific
CI approach applied to a vast number and
various classes of excited states. \cite{Kossoski_2023}
Finally, selected configuration interaction (SCI) methods \cite{Bender_1969,Huron_1973,Buenker_1974,Evangelisti_1983,Angeli_2001c,Liu_2016b,Harrison_1991,Giner_2013,Giner_2015,Holmes_2016,Schriber_2016,Tubman_2016,Sharma_2017,Coe_2018,Garniron_2019,Zhang_2020,Zhang_2021,Mussard_2018,Chien_2018,Loos_2020i,Yao_2020,Damour_2021,Yao_2021,Larsson_2022,Coe_2022,Holmes_2017,Eriksen_2020a,Damour_2023}
are known for their ability to provide results approaching full CI (FCI)
quality, though they are usually restricted to relatively small systems and
basis sets.

Similar studies have been undertaken within the context of CC theory. Notably,
the work of Lee \textit{et al.} showed that the $\Delta$CC approach can provide
accurate excitation energies for both doubly-excited and double-core-hole states. \cite{Lee_2019}
Mayhall and Raghavachari achieved something similar a decade earlier
when they successfully converged multiple coupled-cluster with singles and
doubles (CCSD) solutions for the \ce{NiH} molecule. \cite{Mayhall_2010}
Other works reported the use of $\Delta$CC,\cite{Dreuw_2023,Rishi_2023} notably
for processes implying core excitations. \cite{Zheng_2019,Zheng_2020,Tsuru_2021,Matz_2022,Jayadev_2023,Matz_2023,Arias-Martinez_2022}
This methodology has also been exploited by Schraivogel and Kats within the
distinguishable cluster approximation \cite{Kats_2013,Kats_2014} yielding
promising results in the case of open-shell systems as well as excited states. \cite{Schraivogel_2021}
This kind of approach has been also applied for low-order and cheaper CC methods such as
pair coupled-cluster doubles. \cite{Kossoski_2021,Marie_2021,Rishi_2023}

All these works have been inspired by the discovery of multiple CC solutions.
{\v Z}ivkovi{\'c} and Monkhorst were the first to study the conditions
under which higher roots of the CC equations exist, \cite{Zivkovic_1977,Zivkovic_1978}
while Adamowicz and Bartlett explored the feasibility of reaching certain excited
states of the \ce{LiH} molecule. \cite{Adamowicz_1985} Subsequently,
Jankowski \textit{et al.} clearly evidenced that some of these non-standard
CC solutions are unphysical. \cite{Jankowski_1994,Jankowski_1994a,Jankowski_1995}
A pivotal development in the study of non-standard CC solutions was the
introduction of the homotopy method \cite{Verschelde_1994} by Kowalski,
Jankowski, and others. \cite{Kowalski_1998,Kowalski_1998a,Jankowski_1999,Jankowski_1999a,Jankowski_1999b,Jankowski_1999c,Kowalski_2000,Kowalski_2000a}
We refer the interested reader to the review of Piecuch and Kowalski for
additional information on this topic. \cite{Piecuch_2000}
(See also Refs.~\onlinecite{Csirik_2023a,Csirik_2023b,Faulstich_2022,Faulstich_2023} for a more mathematical perspective.)

Unfortunately, some electronic states require a multireference treatment
to efficiently achieve an accurate description of their electronic structure.
A simple example is the singly-excited state of a closed-shell molecule,
which must be described by at least two determinants related by a global
spin-flip transformation, forming a single configuration state function
(CSF). Within the CI formalism, a multireference treatment is
relatively straightforward as it only requires designing an appropriate
reference before applying excitations. \cite{Szalay_2012,Lischka_2018}
However, on the CC side, the situation is considerably more complex and
implies the use of multireference CC (MRCC) methods. To avoid this, Tuckman
and Neuscamman recently proposed an alternative way to deal with open-shell
singlet excited states within the SRCC formalism. \cite{Tuckman_2023a,Tuckman_2023b}

MRCC methods have been a center of interest since the inception of SRCC
methods, despite the various challenges they present. Interested readers
can find reviews on this topic elsewhere. \cite{Bartlett_2007,Musial_2008,Lyakh_2012,Kohn_2013,Evangelista_2018}
In MRCC, the wave function is constructed through the application of a
wave operator on a model wave function composed of multiple Slater
determinants. MRCC methods can be separated into two main categories,
namely Fock-space MRCC (FS-MRCC),\cite{Lindgren_1978,Haque_1984,Sinha_1986,Pal_1987,Pal_1988,Jeziorski_1989,Chaudhuri_1989,Landau_1999,Landau_2000}
also known as valence-universal MRCC (VU-MRCC), and Hilbert-space MRCC (HS-MRCC).\cite{Jeziorski_1981,Lindgren_1987,Mukherjee_1989,Kucharski_1991b,Balkova_1991a,Balkova_1991b,Balkova_1992,Kucharski_1992,Paldus_1993,Masik_1998,Pittner_1999,Hubac_2000,Pittner_2003}
The primary distinction between these two categories lies in their
respective configuration spaces. FS-MRCC computes states within the
Fock space, i.e., with a variable number of electrons. In contrast,
HS-MRCC considers only determinants within a specified Hilbert space,
i.e., with a fixed number of electrons. Methods based on HS-MRCC can be designed in several
manners. In the state-universal (SU) formalism,\cite{Li_2003,Li_2004}
one considers multiple electronic states, while the single-root or
state-specific formalism looks at each state separately. \cite{Mukherjee_1989,Mahapatra_1998,Masik_1998,Pittner_1999,Hubac_2000,Das_2010,Garniron_2017}
The intermediate Hamiltonian formalism leads to a restricted number
of relevant states among all the computed roots. \cite{Malrieu_1985,Meissner_1995,Meissner_1998,Landau_1999,Landau_2000,Giner_2016,Giner_2017}

One common challenge encountered with multi-state approaches is the potential
emergence of intruder states. However, an interesting observation is that
the open-shell singlet and triplet states that are correctly described by
two determinants can be effectively treated with the SU-MRCC approach using
an incomplete active space (IAS-SU-MRCC) containing only the two determinants
of interest. The resulting scheme is then free of the intruder state problem. Moreover, while
the use of the Jeziorski-Monkhorst ansatz in SU-MRCC usually requires solving
simultaneous equations for all the references, \cite{Jeziorski_1981,Kucharski_1991b}
in this particular case, the equations are considerably simplified due to the
spin-flip relationship between the two reference determinants. It leads to the
so-called two-determinant CC (TD-CC) approach developed by Bartlett's group.\cite{Balkova_1991a,Balkova_1992,Balkova_1993,Balkova_1994,Szalay_1994,Lutz_2018}
Likewise, Piecuch and coworkers have formulated a spin-adapted version of TD-CC. \cite{Piecuch_1992,Piecuch_1993,Piecuch_1994a,Li_1994,Piecuch_1994b,Piecuch_1994c,Piecuch_1995}

In this manuscript, we employ state-specific CCSD and TD-CCSD to
consistently treat various classes of excited states. The vertical
excitation energies are computed within the $\Delta$CCSD framework
and subsequently compared with EOM-CCSD and state-of-the-art high-order
EOM-CC or extrapolated FCI (exFCI) calculations. Both ground-state HF
and state-specific orbitals are employed to build the reference, chosen
as a single- or two-determinant wave function depending on the targeted
state, as discussed in detail below.
Our main goal is to assess the performance of state-specific CC methods to describe different types of excited states.
The present manuscript is organized
as follows. Section \ref{sec:theo} recalls the working equations of SRCC
and MRCC. It also quickly explains how one can compute excitation energies
within the different formalisms.
Section \ref{sec:comp_det} presents the computational details, while Sec.~\ref{sec:res} discusses the main results and compares the performance of the state-specific approach with that of the standard EOM-CCSD method.
Our conclusions are drawn in Sec.~\ref{sec:ccl}.

%%%%%%%%%%%%%%%%%%%%%%%%%%%%%%%%%%%%%%%%%%%%%%%
\section{Theory}
\label{sec:theo}
%%%%%%%%%%%%%%%%%%%%%%%%%%%%%%%%%%%%%%%%%%%%%%%

%==============================================
\subsection{Single-Reference Coupled Cluster}
\label{sec:srcc}
%==============================================

In CC theory, the wave function is expressed as an exponential ansatz
\begin{equation}
  \ket{\Psi} = e^{\hT} \ket{\Phi_0}
\end{equation}
where an excitation operator
\begin{equation}
  \hT = \sum_{n=1}^N \hT_n
\end{equation}
is exponentiated and applied on a reference determinant $\ket{\Phi_0}$
to create new determinants. The excitation operator is composed of a sum
of $n$-electron operators
\begin{equation}
  \hT_n = (n!)^{-2} \sum_{ij\cdots} \sum_{ab\cdots} t_{ij\cdots}^{ab\cdots} \cre{a} \cre{b} \cdots \ani{j} \ani{i}
\end{equation}
associated with cluster amplitudes $t_{ij\cdots}^{ab\cdots}$ that promote
$n$ electrons from occupied spin orbitals $(\phi_i,\phi_j,\hdots)$ in
$\ket{\Phi_0}$, to virtual spin orbitals $(\phi_a,\phi_b,\hdots)$ using
respectively annihilation ($\ani{}$) and creation ($\cre{}$) second-quantized
operators. For the sake of simplicity, we shall use the following shortcut
notations: $\hid{p} \equiv \hid{a}_p$ and $\hi{p} \equiv \hi{a}_p$.

The equations for the energy and the amplitudes are determined using the
time-independent Schr\"odinger equation
\begin{equation}
  \hH \ket{\Psi} = E \ket{\Psi}
\end{equation}
by projecting on either $\bra{\Phi_0} e^{-\hT}$ for the energy
\begin{equation}
  \mel*{\Phi_0}{\bH}{\Phi_0} = E
\end{equation}
or $\bra*{\Phi_{ij\cdots}^{ab\cdots}} e^{-\hT}$ for the amplitudes
\begin{equation}
  \mel*{\Phi_{ij\cdots}^{ab\cdots}}{\bH}{\Phi_0} = 0
\label{eq:t_srcc}
\end{equation}
where $\bra*{\Phi_{ij\cdots}^{ab\cdots}} = \bra*{\Phi_0} \hid{i} \hid{j} \cdots \hi{b} \hi{a}$
is an excited determinant and $\bH = e^{-\hT} \hH e^{\hT}$ is the
(non-Hermitian) similarity-transformed Hamiltonian. To ensure that
the number of unknown amplitudes matches the number of equations,
the latter equation must be solved for all the excited determinants
that can be built from the application of $\hT$ on $\ket{\Phi_0}$.

The second-quantized form of the Hamiltonian operator mentioned in the
previous equations is
\begin{equation}
  \hH = \sum_{pq} \mel*{\phi_p}{\hh}{\phi_q} p^\dag q + \frac{1}{4} \sum_{pqrs} \mel*{\phi_p\phi_q}{}{\phi_r\phi_s} \hid{p} \hid{q}  \hi{s} \hi{r}
\end{equation}
where $p,q,r,s$ are general indices that run over all spin orbitals
$(\phi_i \phi_j \cdots \phi_a \phi_b \cdots)$, $\mel*{\phi_p}{\hh}{\phi_q}$
are the elements of the core Hamiltonian and $\mel*{\phi_p \phi_q}{}{\phi_r \phi_s}$
are the antisymmetrized two-electron integrals. It is convenient to define the
\emph{normal-ordered} Hamiltonian 
\begin{equation}
  \hH_N = \sum_{pq} \mel*{\phi_p}{\hf}{\phi_q} \left\{ \hid{p} \hi{q} \right\} + \frac{1}{4} \sum_{pqrs} \mel*{\phi_p\phi_q}{}{\phi_r\phi_s} \left\{ \hid{p} \hid{q} \hi{s} \hi{r} \right\}
\end{equation}
where $\left\{ \cdots \right\}$ means that the second-quantized operators inside
the curly brakets are normal-ordered with respect to the Fermi vacuum $\Phi_0$
and $\mel*{\phi_p}{\hf}{\phi_q}$ are the elements of the Fock matrix. More
detailed explanations about normal ordering and CC theory can be found
elsewhere. \cite{Crawford_2000,Shavitt_2009}

%==============================================
\subsection{Equation-of-Motion Coupled Cluster}
\label{sec:eomcc}
%==============================================

In CC theory, excited states can be accessed via the EOM \cite{Rowe_1968a,Emrich_1981,Sekino_1984,Geertsen_1989,Stanton_1993a,Comeau_1993,Watts_1994} and LR \cite{Monkhorst_1977,Dalgaard_1983,Sekino_1984,Koch_1990c,Koch_1990a} formalisms.
Both frameworks yield identical excitation energies but result in
different molecular properties. \cite{Bartlett_2007,Sarkar_2021,Chrayteh_2021,Damour_2023} In the
following subsection, we provide a brief overview of EOM-CC to emphasize
the distinctions between this approach and its state-specific counterpart
discussed later.

In EOM-CC, one seeks to solve the following modified Schr\"odinger equation,
\begin{equation}
  \hH_N \ket*{\Psi^{(k)}} = \Delta E^{(k)} \ket*{\Psi^{(k)}}
\label{eq:eom}
\end{equation}
where $\ket*{\Psi^{(k)}}$ is the $k$th excited state of energy $E^{(k)}$ , $\Delta E^{(k)} = E^{(k)} - E_0$, and $E_0$ the energy of the reference wave function.
This excited state is created from the CC ground-state wave function
\begin{equation}
  \ket*{\Psi^{(0)}} = e^{\hT} \ket*{\Phi_0}
\end{equation}
of energy $E^{(0)}$ by applying a linear excitation operator $\hR^{(k)}$, such that
\begin{equation}
  \ket*{\Psi^{(k)}} = \hR^{(k)} \ket*{\Psi^{(0)}}
\end{equation}
with
\begin{gather}
  \hR^{(k)} = r_0 + \sum_{n=1}^N \hR_n^{(k)}
  \\
  \hR_n^{(k)} = (n!)^{-2} \sum_{ij\cdots} \sum_{ab\cdots} r_{ij\cdots}^{ab\cdots} \hid{a} \hid{b} \cdots \hi{j} \hi{i}
\end{gather}
where $r_{ij\cdots}^{ab\cdots}$ are the EOM right amplitudes.
Hence, by multiplying the left-hand side of Eq.~\eqref{eq:eom} by $e^{-\hT}$
and rearranging the equation, one gets
\begin{equation}
  \bH_\text{N} \hR^{(k)} \ket{\Phi_0} = \Delta E^{(k)} \hR^{(k)} \ket{\Phi_0}
\end{equation}
where $\bH_\text{N} = e^{-\hT} \hH_\text{N} e^{\hT}$ is the normal-ordered
version of the similarity-transformed Hamiltonian. \alert{Because the matrix
representation of $\bH_\text{N}$ and $\bH_\text{N}$ share, in the limit 
of a complete basis, the same spectrum of eigenenergies}, the diagonalization
of $\bH_\text{N}$ within a suitable 
basis of Slater determinants yields the energies $\Delta E^{(k)}$
and the corresponding eigenvectors $\hR^{(k)} \ket{\Phi_0}$. Note that,
because $\bH$ is non-hermitian, two sets of biorthogonal eigenvectors
can be obtained, usually referred to as the right and left eigenvectors,
and denoted respectively as $\hR^{(k)} \ket{\Phi_0}$ and $\bra{\Phi_0} \hat{L}^{(k)}$,
respectively, where $\hat{L}^{(k)}$ is a linear de-excitation operator.
Additional details can be found elsewhere. \cite{Crawford_2000,Shavitt_2009}

%==============================================
\subsection{Multi-Reference Coupled Cluster}
\label{sec:mrcc}
%==============================================

Within the MRCC formalism, one seeks a wave operator $\hOmega$, which creates
the exact wave function $\ket{\Psi}$ when applied to a model wave function
$\ket{\Psi_0}$, i.e.,
\begin{equation}
\label{eq:wave_operator}
  \ket{\Psi} = \hOmega \ket{\Psi_0}
\end{equation}
and fulfills the so-called Bloch equation
\begin{equation}
\label{eq:bloch}
  \hH \hOmega = \hOmega \hH \hOmega
\end{equation}
The model wave function, which lives in the model space $\mM_0$, is built
as a linear combination of determinants $\ket{I}$ associated with coefficients
$c_I$, as follows:
\begin{equation}
  \ket{\Psi_0} = \sum_{I \in \mM_0} c_I \ket{I}
\end{equation}
From this, one can define a projector operator onto $\mM_0$
\begin{equation}
  \hP  = \sum_{I \in \mM_0} \hP_I = \sum_{I \in \mM_0} \dyad{I}
\end{equation}
For the sake of simplicity, we assume that $\Psi_0$ is normalized and we
adopt the intermediate normalization between $\Psi_0$ and $\Psi$, i.e.,
$\braket{\Psi_0}{\Psi_0} = 1$ and $\braket{\Psi_0}{\Psi} = 1$.

The key strength of this multireference formalism appears when
Eq.~\eqref{eq:wave_operator} is plugged into the Schr\"odinger equation,
and subsequently multiplied by $\hP$ on the left-hand side, which yields
\begin{equation}
  \hP \hH \hOmega \hP \ket{\Psi_0} = E \ket{\Psi_0}
\end{equation}
because $\hP \hOmega = \hOmega \hP = \hP$ and $\hP \ket{\Psi_0} = \ket{\Psi_0}$. Thus, it can be rewritten as
\begin{equation}
  \hHeff \ket{\Psi_0} = E \ket{\Psi_0}
\end{equation}
meaning that the eigenvalues of $\hH$ can be obtained by only diagonalizing
the effective Hamiltonian
\begin{equation}
\label{eq:Heff}
  \hHeff = \hP \hH \hOmega \hP
\end{equation}
which has the size of the model space $\mM_0$.
As TD-CC is based on HS-MRCC, we briefly review the latter in Sec.~\ref{sec:hs-mrcc}.
The interested reader can find additional details about MRCC (including its
Fock-space variant) elsewhere. \cite{Shavitt_2009,Lyakh_2012}

%==============================================
\subsection{Hilbert-Space MRCC}
\label{sec:hs-mrcc}
%==============================================

In Hilbert-space MRCC, one considers the Jeziorski-Monkhorst form of the
wave operator,\cite{Jeziorski_1981} which reads
\begin{equation}
  \label{eq:omega}
  \hOmega = \sum_{I \in \mM_0} e^{^I\hT} \hP_I
\end{equation}
where $^I\hT$ is a cluster operator defined for the reference $I$
that contains restrictions to remove all the excitations within
the model space.

The amplitude equations related to each determinant within the
model space are obtained by recasting the previous expression into
the Bloch equation \eqref{eq:bloch},
projecting onto a determinant within the model space $\ket{A}$
and a determinant outside the model space $\bra{A_{ij\cdots}^{ab\cdots}}$
\begin{equation}
  \mel*{A_{ij\cdots}^{ab\cdots}}{\hH e^{^A\hT}}{A} - \sum_{I \in \mM_0} \mel*{A_{ij\cdots}^{ab\cdots}}{e^{^I\hT}}{I} \mel*{I}{\hH e^{^A\hT}}{A} = 0
\end{equation}
where $\Heff_{IA} = \mel*{I}{\hH e^{^A\hT}}{A}$ are the matrix
elements of the effective Hamiltonian defined in Eq.~\eqref{eq:Heff}.

These amplitude equations must be solved simultaneously for all the
references $A$ within the model space. As the matrix elements $\Heff_{IA}$
depend on the reference-dependent amplitudes $^A\hT$,
they must be updated, at each iteration, after the computation of the
new amplitudes.
Nonetheless, in the special case where the model space is restricted to
only two determinants related by a global spin-flip transformation, one
ends up with the TD-CC equations that require only one set of amplitudes
to be solved. We explore this special case
in Sec.~\ref{sec:tdcc}.

%==============================================
\subsection{Two-Determinant Coupled Cluster}
\label{sec:tdcc}
%==============================================

\begin{figure}
\includegraphics[width=0.5\linewidth]{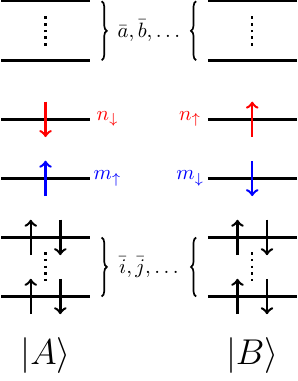}
\caption{Occupancy of the two reference determinants in TD-CC, $\ket{A}$ and $\ket{B}$,
which are related by a global spin-flip transformation.}
\label{fig:tdcc}
\end{figure}

In TD-CC, the reference space is built from two open-shell
determinants, $\ket{A}$ and $\ket{B}$, related by a global
spin-flip transformation:
\begin{align}
  \ket{A} & = \hid{n}_\downarrow \hid{m}_\uparrow \ket{\Phic}
  &
  \ket{B} & = \hid{n}_\uparrow \hid{m}_\downarrow \ket{\Phic}
\label{eq:AB}
\end{align}
where $\Phic$ is a closed-shell determinant containing the common
doubly-occupied orbitals of $A$ and $B$.
These determinants are represented in Fig.~\ref{fig:tdcc}, where
the labels corresponding to the different classes of orbitals are
specified: $\bar{i},\bar{j},\hdots$ for doubly-occupied orbitals,
$m$ or $n$ for active or singly-occupied orbitals, and $\bar{a},\bar{b},\hdots$
for virtual orbitals.
Thus, in the following, occupied and virtual spin orbitals are
labeled $i,j,\hdots$ and $a,b,\hdots$, respectively. For the
reference $\ket{A}$, we have
\begin{subequations}
\begin{align}
    \qty{i,j,\hdots} &= \qty{\bar{i},\bar{j},\hdots} \cup \qty{m_\uparrow,n_\downarrow}
    \label{eq:occ_A}
    \\
    \qty{a,b,\hdots} &= \qty{\bar{a},\bar{b},\hdots} \cup \qty{m_\downarrow,n_\uparrow}
    \label{eq:vir_A}
\end{align}
\end{subequations}
and likewise for the reference $\ket{B}$
\begin{subequations}
\begin{align}
    \qty{i,j,\hdots} &= \qty{\bar{i},\bar{j},\hdots} \cup \qty{m_\downarrow,n_\uparrow}
    \label{eq:occ_B}
    \\
    \qty{a,b,\hdots} &= \qty{\bar{a},\bar{b},\hdots} \cup \qty{m_\uparrow,n_\downarrow}
    \label{eq:vir_B}
\end{align}
\end{subequations}

%%% FIG 1 %%%
%%% %%% %%%

The model space is thus composed
of one singlet (S) and one triplet (T) model wave function:
\begin{align}
  \ket{^\text{S}\Psi_0} & = \frac{\ket{A} + \ket{B}}{\sqrt{2}}
  &
  \ket{^\text{T}\Psi_0} & = \frac{\ket{A} - \ket{B}}{\sqrt{2}}
\end{align}
Thus, using Eq.~\eqref{eq:wave_operator} and Eq.~\eqref{eq:omega}, the TDCC singlet and triplet wave functions are
\begin{align}
  \ket{^\text{S}\Psi} & = \frac{e^{^AT}\ket{A} + e^{^BT}\ket{B}}{\sqrt{2}}
  &
  \ket{^\text{T}\Psi} & = \frac{e^{^AT}\ket{A} - e^{^BT}\ket{B}}{\sqrt{2}}
\end{align}

Because of the spin-flip relation between these model wave functions,
the diagonalization of the effective Hamiltonian leads to
the following energies for the singlet and triplet states:
\begin{subequations}
\begin{align}
  \ES & = \mel*{A}{\hH_\text{N} e^{{}^A\hT}}{A} + \mel*{A}{\hH_\text{N} e^{{}^B\hT}}{B}
  \\
  \ET & = \mel*{A}{\hH_\text{N} e^{{}^A\hT}}{A} - \mel*{A}{\hH_\text{N} e^{{}^B\hT}}{B}
\end{align}
\end{subequations}
At the TD-CCSD level, the excitation operators are restricted to single
and double excitations with respect to their respective references.
For example, we have
\begin{equation}
\begin{split}
  {}^A\hT
  & = {}^A\hT_1 + {}^A\hT_2
  \\
  & = \sum_{ia} {}^At_{i}^{a} \hid{a} \hi{i}
  + \frac{1}{4} \sum_{ijab} {}^At_{ij}^{ab} \hid{a} \hid{b} \hi{j} \hi{i}
\end{split}
\end{equation}
with a similar expression for the reference $B$.
Note that excitations of the form
$t_{{m_\uparrow}{n_\downarrow}}^{{m_\downarrow}{n_\uparrow}} \hid{m}_\downarrow \hid{n}_\uparrow \hi{n}_\downarrow \hi{m}_\uparrow $, which link $\ket{A}$ to $\ket{B}$,
are removed from the summation since the excitations between two determinants
of the active space must be discarded.

The ${}^AT_1$ equations for the reference-dependent amplitudes are
\begin{equation}
\label{eq:t1_tdcc}
\begin{split}
  {}^AQ_i^a
  & = \mel*{A_{i}^{a}}{\hH_N e^{{}^A\hT}}{A}_\text{c} - \qty( \mel*{A_i^a}{e^{{}^B\hT}}{B} \Heff_{BA} )_\text{c}
  \\
  & = \mel*{A_{i}^{a}}{\hH_N e^{{}^A\hT}}{A}_\text{c} - {{}^A}M_i^a \Heff_{BA} = 0
\end{split}
\end{equation}
while, for the ${}^AT_2$ equations, we have
\begin{equation}
\label{eq:t2_tdcc}
\begin{split}
  {}^AQ_{ij}^{ab} &= \mel*{A_{ij}^{ab}}{\hH_N e^{{}^A\hT}}{A}_\text{c}
  - \qty( \mel*{A_{ij}^{ab}}{e^{{}^B\hT}}{B} \Heff_{BA} )_\text{c}
  \\
  & - P(ij,ab) \qty[ \qty( {{}^A}t_i^a - \hat{R}_{ia} {}^Bt_i^a ) \qty( \mel*{A_j^b}{e^{{}^B\hT}}{B} \Heff_{BA} )_\text{c} ]
  \\
  & = \mel*{A_{ij}^{ab}}{\hH_N e^{{}^A\hT}}{A}_\text{c} - {{}^A}M_{ij}^{ab} \Heff_{BA} = 0
\end{split}
\end{equation}
where $\hH_N$ is normal-ordered with respect to $A$,
$M_{i}^{a}$, and $M_{ij}^{ab}$ are the so-called renormalization terms,
the subscript $\text{c}$ indicates that only connected terms are considered,
$\hat{R}_{ia} = (1 - \delta_{in_\uparrow}) (1 - \delta_{im_\downarrow}) (1 - \delta_{am_\uparrow}) (1 - \delta_{an_\downarrow})$, and $P(ij,ab) = 1 - (i \leftrightarrow j) - (a \leftrightarrow b) + (i \leftrightarrow j) (a \leftrightarrow b)$ is a permutation operator.
It is worth noting that one only needs to solve Eqs.~\eqref{eq:t1_tdcc}
and \eqref{eq:t2_tdcc} for the reference $A$ since the corresponding
amplitudes for $B$ can be easily obtained thanks to the global spin-flip
relationship between $A$ and $B$.
Further explanations and explicit expressions of each quantity can be
found in Refs.~\onlinecite{Balkova_1992,Szalay_1994,Lutz_2018}.

Furthermore, as explained by Lutz \textit{et al.}, \cite{Lutz_2018} only
two out of four determinants contribute to the singlet and triplet states
in the case of a complete active space (CAS) containing two electrons in two
orbitals of different spatial symmetries (as for \ce{H2} in a minimal basis set).
Therefore, TD-CC is performed on top of a (2,2) complete active space, and the CC
single excitation amplitudes between the active orbitals are zero,
further simplifying the equations. However, in the case where the active
space contains two orbitals of the same spatial symmetry,
TD-CC accounts for an incomplete active space.
The working equations for both scenarios are presented in the \SupMat.

%==============================================
\subsection{Derivation of the TD-CCSD Equations}
\label{sec:formula_tdcc}
%==============================================

The TD-CCSD equations can be derived by hand using Wick's theorem or
diagrammatic techniques.\cite{Kucharski_1991b}
However, the manual manipulation of a substantial
number of equations and indices is highly prone to errors,
especially during their implementation in computer software.
Fortunately, software solutions are available to streamline this process.
Notably, tools such as \texttt{p$^\dag$q} by Rubin and DePrince \cite{Rubin_2021}
and \textsc{wick\&d} developed by Evangelista \cite{Evangelista_2022} can
algebraically derive this kind of equations in various contexts.

Let us have a look at Eq.~\eqref{eq:t1_tdcc} for the $^AT_1$ amplitudes.
The first term in the right-hand-side of Eq.~\eqref{eq:t1_tdcc},
$\mel*{A_{i}^{a}}{H_N e^{{}^A\hT}}{A}_\text{c}$, corresponds to the
standard term of the $T_1$ SRCC equations considered
in Eq.~\eqref{eq:t_srcc}, with a restriction over the summations
in $^A\hT$. The only new part is the renormalization term,
$\qty( \mel*{A_i^a}{e^{{}^B\hT}}{B} \Heff_{BA} )_\text{c}$ .
To compute the latter term, it is convenient to express each
component with respect to a unique reference. For example, the term
$\Heff_{BA} = \mel*{B}{\hH e^{^A\hT}}{A}$
can be obtained by rewriting $\ket{B}$ as an excitation from $\ket{A}$
using Eq.~\eqref{eq:AB}, i.e.,
\begin{equation}
\begin{split}
    \ket{B}
    & = \hid{n}_\uparrow \hid{m}_\downarrow \ket{\Phic}
    \\
    & = \hid{n}_\uparrow \hid{m}_\downarrow \hi{m}_\uparrow \hi{n}_\downarrow \hid{n}_\downarrow \hid{m}_\uparrow \ket{\Phic}
    \\
    & = \hid{n}_\uparrow \hid{m}_\downarrow \hi{m}_\uparrow \hi{n}_\downarrow \ket{A}
    \\
    & = \ket{A_{{n_\downarrow} {m_\uparrow}}^{{n_\uparrow} {m_\downarrow}}}
\end{split}
\end{equation}
Thus, the quantity that needs to be evaluated is
$\mel*{A_{{n_\downarrow} {m_\uparrow}}^{{n_\uparrow} {m_\downarrow}}}{\hH e^{^A\hT}}{A}$,
which is simply one of the SRCC $T_2$ equations with additional restrictions
over the summations in $^A\hT$.

The derivation of the second part of the renormalization term, $M_a^i = \mel*{A_i^a}{e^{{}^B\hT}}{B}$,
is a little bit more involved. Due to the presence of $e^{{}^B\hT}\ket{B}$,
it is easier to express $\bra{A_i^a}$ as an excitation from the reference $\bra{B}$.
To do so, we start by rewriting $\ket{A}$ as an excitation from $\ket{B}$:
\begin{equation}
\begin{split}
  \ket{A}
  & = \hid{n}_\downarrow \hid{m}_\uparrow \ket{\Phic}
  \\
  & = \hid{n}_\downarrow \hid{m}_\uparrow \hi{m}_\downarrow \hi{n}_\uparrow \hid{n}_\uparrow \hid{m}_\downarrow \ket{\Phic}
  \\
  & = \hid{n}_\downarrow \hid{m}_\uparrow \hi{m}_\downarrow \hi{n}_\uparrow \ket{B}
\end{split}
\end{equation}
which yields
\begin{equation}
  \ket{A_{i}^{a}} = \hat{a}^\dag \hat{i} \hid{n}_\downarrow \hid{m}_\uparrow \hi{m}_\downarrow \hi{n}_\uparrow \ket{B}
\end{equation}

At this stage, a difficulty arises from the fact that $A$ and $B$
do not share the same set of occupied and virtual spin orbitals
because of the two unpaired
electrons, as shown in Fig.~\ref{fig:tdcc} and Eqs.~\eqref{eq:occ_A}, \eqref{eq:occ_B}, \eqref{eq:vir_A}, and \eqref{eq:vir_B}.
To solve this issue, we generate a set of strings of
second-quantized operators corresponding to every possible
combination of occupancy and spin cases, and apply them to $\bra{B}$:
\begin{multline}
  \label{eq:strings}
  \qty(\hat{a}^\dag \hat{i} \hid{n}_\downarrow \hid{m}_\uparrow \hi{m}_\downarrow \hi{n}_\uparrow )^\dag
  \\
  \quad \rightarrow
  \Bigg\{
  \qty(\hid{a}_o \hi{i}_o \hid{n}_\downarrow \hid{m}_\uparrow \hi{m}_\downarrow \hi{n}_\uparrow )^\dag,
  \qty(\hid{a}_v \hi{i}_o \hid{n}_\downarrow \hid{m}_\uparrow \hi{m}_\downarrow \hi{n}_\uparrow )^\dag,
  \\
  \qty(\hid{a}_o \hi{i}_v \hid{n}_\downarrow \hid{m}_\uparrow \hi{m}_\downarrow \hi{n}_\uparrow )^\dag,
  \qty(\hid{a}_v \hi{i}_v \hid{n}_\downarrow \hid{m}_\uparrow \hi{m}_\downarrow \hi{n}_\uparrow )^\dag
  \Bigg\}
\end{multline}
where the subscripts $o$ (occupied) and $v$ (virtual) specify
the occupancy of the spin orbitals of reference $A$ in the reference $B$.
Note that we have not specified the spin of $\phi_a$ and $\phi_i$ on purpose
to reduce the number of terms. Otherwise, by including their spin, each string
would lead to four additional terms, that is, $\uparrow\uparrow$, $\uparrow\downarrow$, $\downarrow\uparrow$,
and $\downarrow\downarrow$.
Subsequently, Wick's theorem is applied to each of the generated strings. The
result is a set of normal-ordered strings of second-quantized operators with
Kronecker deltas.

To avoid errors, we have implemented Wick's theorem for a single
string of second-quantized operators in \texttt{Python}.
The source code is freely available on \textsc{github} at \url{https://github.com/LCPQ/SimpleWick}, and archived in the \textsc{software heritage} database. \cite{simplewick}
As a reminder, Wick's theorem is a technique to rewrite an arbitrary string
of second-quantized operators into a sum of normal-order strings. \cite{Wick_1950,Crawford_2000,Shavitt_2009}
Thus, to evaluate
\begin{equation}
\begin{split}
  ^AM_i^a
  & = \mel*{A_i^a}{e^{{}^B\hT}}{B}
  \\
  & = \mel*{B}{(\hat{a}^\dag \hat{i} \hid{n}_\downarrow \hid{m}_\uparrow \hi{m}_\downarrow \hi{n}_\uparrow )^\dag e^{{}^B\hT}}{B}
\end{split}
\end{equation}
the string $(\hat{a}^\dag \hat{i} \hid{n}_\downarrow \hid{m}_\uparrow \hi{m}_\downarrow \hi{n}_\uparrow )^\dag$
is expanded and evaluated as explained above.
The resulting strings of normal-ordered
second-quantized operators are reordered and fed to
\texttt{p$^\dag$q} as left operators of the form $\qty{\hid{i} \hid{j} \cdots \hi{b} \hi{a}}$,
using the occupancy definition of Eq.~\eqref{eq:occ_B} and ~\eqref{eq:vir_B}, to be separately evaluated.
Schematically, we evaluate terms of the form
\begin{equation}
  \mel*{B}{\qty{\hid{i} \hid{j} \cdots \hi{b} \hi{a}} \ e^{{}^B\hT}}{B}
\end{equation}
The use of Wick's theorem allows us to handle the different occupancy cases.
To illustrate this, let us consider
the first term of Eq.~\eqref{eq:strings} that contributes to $^AM_{i_\uparrow}^{a_\uparrow}$.
In this case, we have
\begin{equation}
\begin{split}
\hid{n}_\uparrow \hid{m}_\downarrow \hi{m}_\uparrow \hi{n}_\downarrow \hid{i}_{o\uparrow} \hi{a}_{o\uparrow}
&= \qty{\hid{n}_\uparrow \hid{m}_\downarrow \hi{m}_\uparrow \hi{n}_\downarrow \hid{i}_{o\uparrow} \hi{a}_{o\uparrow}} \\
&+ \delta_{na} \qty{\hid{m}_\downarrow \hi{m}_\uparrow \hi{n}_\downarrow \hid{i}_{o\uparrow}}
\end{split}
\end{equation}
Obviously, the contraction between $a$ and $i$, as well as the contraction between
$n$ and $m$, are zero since they correspond to distinct spin orbitals.
In addition, it is straightforward to see that the first term of the previous equation
yields a zero contribution. Thus, we have
\begin{equation}
^AM_{i_\uparrow}^{a_\uparrow} = \delta_{na} \mel*{B}{\qty{\hid{m}_\downarrow  \hid{i}_{o\uparrow}\hi{m}_\uparrow \hi{n}_\downarrow} \ e^{{}^B\hT}}{B}
\end{equation}
which becomes
\begin{equation}
\begin{split}
^AM_{i_\uparrow}^{n_\uparrow} &= {}^Bt_{m_\downarrow i_\uparrow}^{n_\downarrow m_\uparrow}
+ {}^Bt_{m_\downarrow}^{n_\downarrow} {^B}t_{i_\uparrow}^{m_\uparrow} \\
&= {}^At_{m_\uparrow i_\downarrow}^{n_\uparrow m_\downarrow}
+ {}^At_{m_\uparrow}^{n_\uparrow} {^A}t_{i_\downarrow}^{m_\downarrow}
\end{split}
\end{equation}

However, because our aim is to evaluate $\qty( \mel*{A_i^a}{e^{{}^B\hT}}{B} \Heff_{BA} )_\text{c}$,
we must remove the disconnected contributions. To do so,
the terms involving at least one cluster amplitude not connected
to the effective Hamiltonian, i.e., a cluster amplitude that does not contain
any active label, like $m$ or $n$, are excluded.
Then, the generated terms are associated with the
Kronecker deltas and the signs coming from Wick's theorem.
At this stage, all quantities are related to the
reference $B$. We thus perform a global spin-flip transformation on $M_i^a$ to express
the latter quantity with respect to the reference $A$.
Finally, the contributions to $M_i^a$ are generated for computer implementation.
A similar procedure is applied to the renormalization term of the
${}^AT_2$ equations [see Eq.~\eqref{eq:t2_tdcc}].

To perform all these tasks, we have developed a \texttt{Python} code that interfaces \texttt{p$^\dag$q}
with the Wick theorem code described above. It enables to generate {\LaTeX} formulas and
\texttt{Fortran} code with ease within a \texttt{jupyter} notebook. This code is also freely
available on \textsc{github} at \url{https://github.com/LCPQ/Gen_Eq_TDCC}, and archived in the \textsc{software heritage} database.\cite{geneq}
For the sake of completeness, the automatically derived formulas related to Eqs.~\eqref{eq:t1_tdcc} and \eqref{eq:t2_tdcc} can be found in the \SupMat.

%%%%%%%%%%%%%%%%%%%%%%%%%%%%%%%%%%%%%%%%%%%%%%
\section{Computational Details}
\label{sec:comp_det}
%%%%%%%%%%%%%%%%%%%%%%%%%%%%%%%%%%%%%%%%%%%%%%

Most of the molecules and states considered in this paper have been
studied in Ref.~\onlinecite{Kossoski_2023} and were originally extracted
from the \textsc{quest} database. \cite{Veril_2021} The geometries are
also taken from the \textsc{quest} database while the state-specific
orbitals are the same as those of Ref.~\onlinecite{Kossoski_2023} (see below).
We employ the aug-cc-pVDZ basis set for molecules with three
or fewer non-hydrogen atoms and the 6-31+G(d) basis otherwise,
applying the frozen core approximation (large core for third-row atoms).
In the special case of the \ce{Be} atom, the aug-cc-pVDZ basis set was chosen
so as to comply with Ref.~\onlinecite{Loos_2020z}.
Furthermore, we rely
on the same reference values as in Ref.~\onlinecite{Kossoski_2023}, which
are mainly of CC with singles, doubles, triples, and quadruples (CCSDTQ) or exFCI quality.
We consider two new systems, borole and oxalyl fluoride, with geometries optimized at the CC3/aug-cc-pVTZ level of theory following the \textsc{quest} methodology. These are reported in the \SupMat.

Our set of excited states comprises 125 singlet, 36 doublet, and 106
triplet states in small- and medium-sized organic compounds, among which there are 9
doubly-excited states. Compared to Ref.~\onlinecite{Kossoski_2023}, a few states were
removed for reasons detailed below.

In this paper, all the calculations were performed with the \textsc{quantum package}
software, \cite{Garniron_2019} where CCSD (for both open- and closed-shell systems)
and TD-CCSD have been implemented. The calculations were carried out using
the $\Delta$CC strategy. Hence, the excitation energies from the ground
state were straightforwardly determined as an energy difference between
ground and excited states. The ground-state energies were computed with
CCSD and ground-state HF orbitals. Different strategies were applied to
compute excited states based on their nature. For doubly-excited states,
a CCSD calculation was performed on top of a closed-shell doubly-excited
determinant. Similarly, for the doublet excited states, a CCSD calculation was
performed on top of a singly-excited determinant. Finally, for open-shell
singlet and triplet excited states, a TD-CCSD calculation was undertaken
using a pair of open-shell singly-excited determinants as the reference.

Concerning the choice of the orbitals, we employ both the ground-state HF
and state-specific optimized orbitals. At the HF level, we rely systematically
on the restricted HF and restricted open-shell HF (ROHF) formalisms for
closed- and open-shell systems, respectively. The state-specific orbitals,
extracted from our recent work, \cite{Kossoski_2023} were
obtained by optimizing the orbitals for a minimal CSF space using the Newton-Raphson
method, also implemented in \textsc{quantum package}. \cite{Garniron_2019,Damour_2021}
We refer to \Dhf as a $\Delta$CCSD calculation where
ground-state HF orbitals have been considered for both ground and excited states, while \Doo
denotes a $\Delta$CCSD calculation that employs HF orbitals for the
ground state and state-specific optimized orbitals for the excited state.
For \Dhf, the excited-state calculations were performed by changing
the orbital occupancy to match with the excited-state dominant determinant
obtained from a prior EOM-CCSD calculation.
For \Doo, the excited-state
calculations were done for a non-Aufbau single CSF reference for which the orbitals were optimized.
It is important to note that the orbitals in \Doo are optimized for a single CSF wave function.
In other words, they do not minimize the CCSD energy as in orbital-optimized CCSD. \cite{Sherrill_1998,Krylov_1998,Krylov_2000,Bozkaya_2011}

In the statistical analysis presented below, we report the usual indicators:
the mean signed error (MSE), the mean absolute error (MAE), the root-mean-square
error (RMSE), and the standard deviation of the errors (SDE). For the sake of
completeness, the raw data associated with each figure and table, in addition to
the total and excitation energies computed at the different levels of theory are reported in the \SupMat.

%%%%%%%%%%%%%%%%%%%%%%%%%%%%%%%%%%%%%%%%%%%%%%
\section{Results and discussion}
\label{sec:res}
%%%%%%%%%%%%%%%%%%%%%%%%%%%%%%%%%%%%%%%%%%%%%%
%%%%%%%%%%%%%%%%%%%%%%%%%%%%%%%%%%%%%%%%%%%%%%
\subsection{Doubly-excited states}
\label{sec:double}
%%%%%%%%%%%%%%%%%%%%%%%%%%%%%%%%%%%%%%%%%%%%%%

For the doubly-excited states, we limited our focus to a small subset of 9 molecules,
including the 6 listed in Ref.~\citenum{Kossoski_2023} complemented with
borole, oxalyl fluoride, and tetrazine.
Other molecules such as \ce{C2}
and \ce{C3} are excluded from this study since they require consideration
of more than one closed-shell determinant to accurately describe their
doubly-excited states. \cite{Kossoski_2023} This requirement becomes evident when examining
the coefficients of their CI wave function.
%In such cases, the use of $\Delta$CCSD often leads to an error of the order of \SI{1}{\eV}.
Additionally, we restrict ourselves to ``genuine'' double excitations only,
which have a small contribution from the single excitations (see the $\%T_1$ values in Table \ref{tab:doubly})
and are hence reasonably well described by a single closed-shell determinant.
Transitions with a dominant single excitation character, yet a significant 
partial double excitation contribution (large $\%T_1$ values),
such as the famous $2 {}^1A_g$ dark state of butadiene,
\cite{Saha_2006,Watson_2012,Shu_2017,Barca_2018b,Loos_2019,doCasal_2023}
are treated as single excitations using TD-CCSD (see Sec.~\ref{sec:single}).
Most of the reference data for these transitions have been extracted from Ref.~\onlinecite{Loos_2019}.

\begin{figure}[h!]
    \centering
    \includegraphics[width=\linewidth]{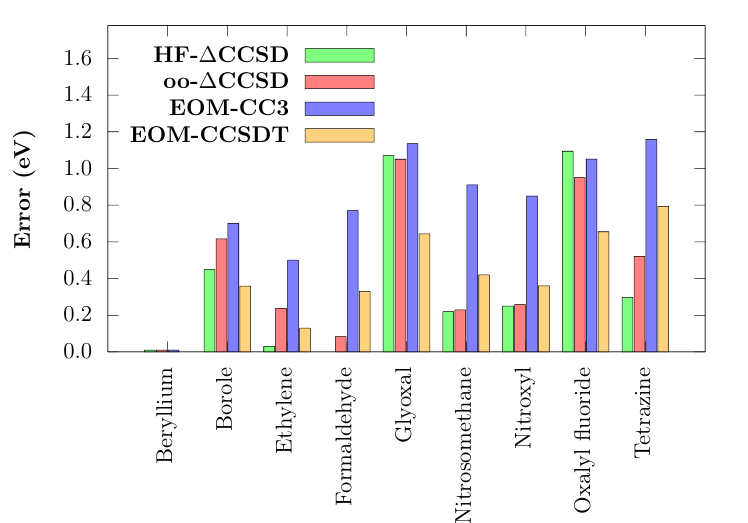}
	\caption{Errors in the excitation energies (with respect to the reference values) of doubly-excited states, computed at various levels of theory. See Table \ref{tab:doubly} for the raw data.}
    \label{fig:doubly}
\end{figure}

The results for each system are represented in Fig.~\ref{fig:doubly} and are also collected in Table
\ref{tab:doubly}, where we further report the $\Delta$SCF results and the value of
$\%T_1$ computed at the CC3/aug-cc-pVTZ level. We do not
consider EOM-CCSD due to the very poor quality of the results. Indeed, it is well
known that, at this level of theory, some of the doubly-excited states are overshooted by several eV.
For the doubly-excited states, we do not perform a statistical analysis due to the limited number of states.

%%% TABLE I %%%
\begin{table*}
	\caption{Excitation energies associated with the doubly-excited states calculated at various levels of theory.
	The error with respect to the reference value is reported in parentheses.}
	\label{tab:doubly}
	\begin{ruledtabular}
		\begin{tabular}{lddrrrrr}
            &  &	\multicolumn{6}{c}{Excitation energies \si{\eV}}\\
            \cline{3-8}
            Molecule &	\tabc{$\%T_1$\fnm[1]} 	&	\tabc{Ref.} & \tabc{$\Delta$SCF} & \tabc{\Dhf\fnm[2]} & \tabc{\Doo} & \tabc{EOM-CC3} & \tabc{EOM-CCSDT}    \\
            \hline
            Beryllium       & 29 &  7.22 &  7.86(+0.64) &  7.23(+0.01) &  7.23(+0.01) &  7.23(+0.01) &  7.22(+0.00) \\
            Borole          & 20 &  4.71 &  5.70(+0.99) &  5.16(+0.45) &  5.33(+0.62) &  5.41(+0.70) &  5.07(+0.36) \\
            Ethylene        & 20 & 13.07 & 13.54(+0.47) & 13.10(+0.03) & 13.31(+0.24) & 13.57(+0.50) & 13.20(+0.13) \\
            Formaldehyde    &  5 & 10.45 & 10.83(+0.38) & 	           & 10.53(+0.08) & 11.22(+0.77) & 10.78(+0.33) \\
            Glyoxal         &  0 &  5.60 & 10.50(+4.90) &  6.67(+1.07) &  6.55(+0.95) &  6.74(+1.14) &  6.24(+0.64) \\
            Nitrosomethane  &  3 &  4.84 &  4.96(+0.12) &  5.06(+0.22) &  5.07(+0.23) &  5.75(+0.91) &  5.26(+0.42) \\
            Nitroxyl        &  0 &  4.40 &  4.57(+0.17) &  4.65(+0.25) &  4.66(+0.26) &  5.25(+0.85) &  4.76(+0.36) \\
            Oxalyl fluoride &  2 &  9.21 & 12.57(+3.37) & 10.30(+1.09) & 10.16(+0.95) & 10.26(+1.05) &  9.86(+0.66) \\
            Tetrazine       &  1 &  5.06 &  7.77(+2.71) &  5.36(+0.30) &  5.58(+0.52) &  6.22(+1.16) &  5.86(+0.79) \\
		\end{tabular}
	\end{ruledtabular}
    \fnt[1]{Percentage of single excitations involved in the transition computed at the CC3/aug-cc-pVTZ level.}
    \fnt[2]{For formaldehyde, we were unable to locate unambiguously the electronic state of interest at this level of theory.}
\end{table*}

Interestingly, we found that \Dhf generally yields similar results to those of \Doo,
with the exception of formaldehyde, for which we were not able to locate the state of interest
with \Dhf.
This result reveals a limited effect of mean-field orbital optimization for $\Delta$CCSD calculations.
It is also worth noticing that both \Dhf and \Doo systematically overestimate the excitation energies,
hinting at the importance of the missing contributions from higher excitations (see below).

Glyoxal and oxalyl fluoride exhibit the largest errors of our set (around \SI{1}{\eV}), with both \Dhf and \Doo.
Based on SCI calculations, for both molecules and with both methods, we found that the CI coefficient
associated with the reference determinant of the excited state is large ($\approx 0.8$).
One possible explanation for the observed discrepancy lies in the existence of a second doubly-excited determinant, relative to both the HF and the reference determinant, which carries a substantial weight in the wave function ($\approx 0.2$).
Indeed, for glyoxal, including this second determinant in the reference space leads to improved results in state-specific CI with singles and doubles ($\Delta$CISD) calculations. \cite{Kossoski_2023}
It is also interesting to notice that the presence of singly-excited determinants with somewhat larger coefficients (up to $\approx 0.3$)
is problematic, based on the smaller errors we observe in the other doubly-excited states. Therefore, the
inaccuracies for glyoxal and oxalyl fluoride may originate from the fact that important single and double excitations
accessed from the second doubly-excited determinant are not accounted for, which would appear as triple and quadruple excitations
with respect to the reference determinant, and are thus lacking in the $\Delta$CCSD calculations.
This also explains the systematically overestimated excitation energies.

When compared to standard EOM-CC methods, the $\Delta$CCSD approach demonstrates its
superiority for the doubly-excited states surveyed here, outperforming EOM-CC3 and even
EOM-CCSDT in some cases. However, $\Delta$CCSD
becomes less accurate when dealing with systems possessing several doubly-excited
determinants with significant weights, as discussed above for glyoxal and oxalyl fluoride.

%%%%%%%%%%%%%%%%%%%%%%%%%%%%%%%%%%%%%%%%%%%%%%
\subsection{Doublet-doublet transitions}
\label{sec:doublet}
%%%%%%%%%%%%%%%%%%%%%%%%%%%%%%%%%%%%%%%%%%%%%%

We included all the
radical doublets from Ref.~\onlinecite{Kossoski_2023}, except for 11 states (outlined below), making a total of 36 states in the present set. The reference data for these molecules have been extracted from Ref.~\onlinecite{Loos_2020z}.
We did not consider the four states of nitromethyl, since their
assignment is particularly challenging due to the significant deviations observed between EOM-CCSD
and EOM-CCSDT. We also excluded one state of \ce{BeH}
as the two states of interest share the same dominant determinant and one state of \ce{BeF},
for which we did not manage to converge the orbital optimization.
Moreover, \Dhf
leads to very large errors for two states of \ce{CH}, as they need two open-shell determinants to be qualitatively described. \cite{Kossoski_2023}
Hence, they have also been discarded.
Finally, we did not consider one of the states of vinyl and the two states of \ce{OH} which need three determinants with three unpaired electrons to be described. \cite{Kossoski_2023}
The detailed results for all the states can be found in the \SupMat.

\begin{figure*}
	\includegraphics[width=0.32\textwidth]{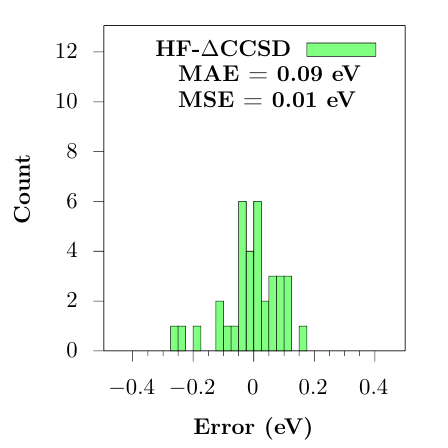}
	\includegraphics[width=0.32\textwidth]{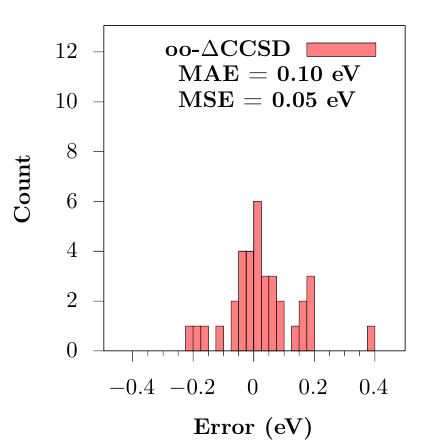}
	\includegraphics[width=0.32\textwidth]{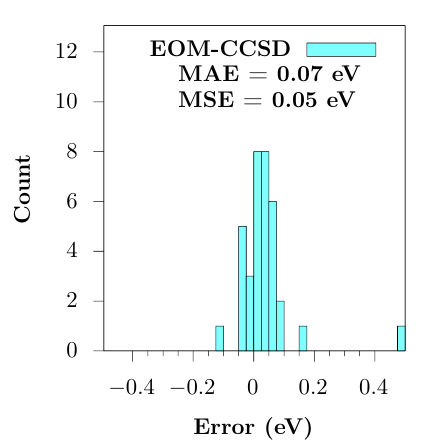}
	\caption{Histograms of the error in excitation energies (with respect to the reference values) computed at the \Dhf (left), \Doo (center), and \eom (right) levels for doublet-doublet transitions.}
	\label{fig:doublet}
\end{figure*}

Our set of 36 doublet-doublet transitions should be large enough to render reasonable statistics.
The distribution of errors with respect to accurate reference values, as obtained with \eom, \Dhf, and \Doo,
is illustrated in Fig.~\ref{fig:doublet}, whereas the associated statistical
measures are reported in Table \ref{tab:doublet}.
When compared to \eom, the distribution of errors for \Dhf and \Doo appears to be less sharp, leading to slightly worse statistics.
Indeed, the distributions are similar for \Dhf and \Doo, with comparable MAEs (\SIrange{0.09}{0.10}{\eV}),
both underperforming \eom, which has a MAE of \SI{0.07}{\eV}. The same can be concluded from the RMSEs and SDEs.
Surprisingly, in contrast to \Dhf, which has a near-zero MSE, \Doo has the same slightly
positive MSE (\SI{0.05}{\eV}) as \eom. Thus, the use of optimized orbitals instead of
HF ground-state orbitals does not improve the overall accuracy. On the contrary, the MAE,
RMSE, and SDE are slightly larger (by \SI{0.01}{\eV} or \SI{0.02}{\eV}) when optimized orbitals are employed, which could be a statistical artifact though.
Overall, the accuracy of $\Delta$CCSD for doublet-doublet transitions
is only marginally worse compared to \eom. 
%\yann{It could be interesting to look at the repartition of the error of $\Delta$CCSD
%depending on the weight of the chosen CSF in \eom. However, due to the limited number
%of states it is hard to draw conclusion on this aspect.}

We further notice that \Doo is more accurate than the $\Delta$CISD calculations reported in Ref.~\onlinecite{Kossoski_2023},
with respective MAEs of \SI{0.10}{\eV} and \SI{0.16}{\eV}.
Since both methods employ the same set of state-specific orbitals, reference wave function, and excitation manifold (singles and doubles),
the superior performance of \Doo can be attributed to the inclusion of higher-order excitations accessed through the connected terms of CC.

%%% TABLE II %%%
\begin{table}
   \caption{Statistical indicators associated with the errors in
doublet-doublet transitions computed at the \eom, \Dhf, and \Doo levels.
All values are in \si{\eV}.}
   \label{tab:doublet}
   \begin{ruledtabular}
       \begin{tabular}{lddddd}
       Method   & \tabc{\# states}  & \tabc{MAE}  &  \tabc{MSE}  & \tabc{RMSE} & \tabc{SDE} \\
            \hline
            EOM-CCSD &     36 &   0.07 &   0.05 &   0.13 &   0.12 \\
            \Dhf     &     36 &   0.09 &   0.01 &   0.15 &   0.15 \\
            \Doo     &     36 &   0.10 &   0.05 &   0.17 &   0.16
       \end{tabular}
   \end{ruledtabular}
\end{table}

%%%%%%%%%%%%%%%%%%%%%%%%%%%%%%%%%%%%%%%%%%%%%%
\subsection{Open-shell singly-excited states}
\label{sec:single}
%%%%%%%%%%%%%%%%%%%%%%%%%%%%%%%%%%%%%%%%%%%%%%

Let us now focus on the performance of $\Delta$CCSD for open-shell singly-excited
states of closed-shell systems. Out of the 237 states considered, two could not be targeted with our protocol for \Dhf calculations, since their
dominant determinants are the same as those of a lower-lying
excited state. In addition, HF-TDCC calculations did not converge for four other states,
making a total of 231 converged states. All the results are available in the \SupMat, with the above-mentioned problematic cases highlighted in magenta.

\begin{figure*}
	\includegraphics[width=0.32\textwidth]{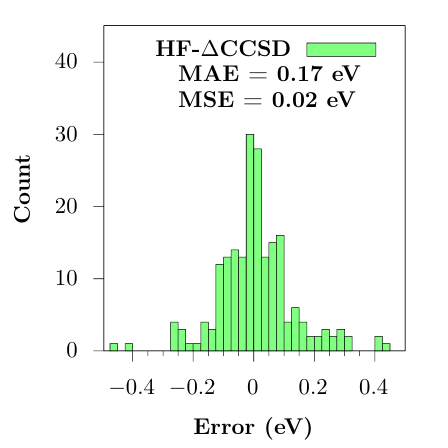}
	\includegraphics[width=0.32\textwidth]{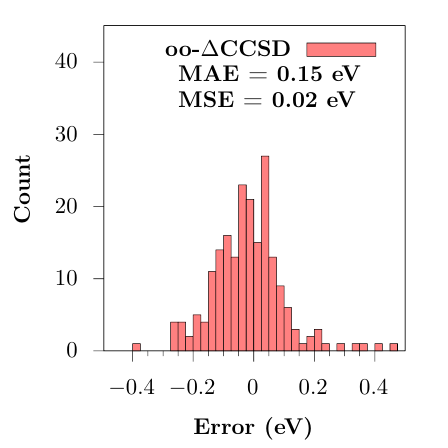}
	\includegraphics[width=0.32\textwidth]{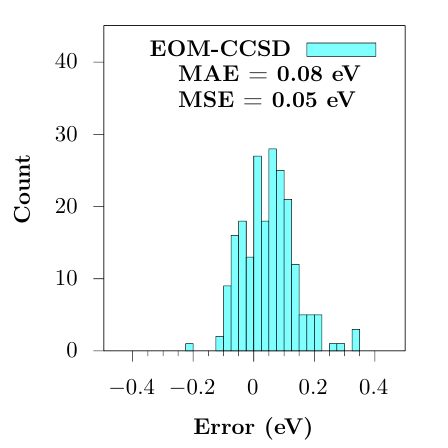}
	\caption{Histograms of the error in excitation energies (with respect to the reference values) at the \Dhf (left), \Doo (center), and \eom (right) levels, for open-shell singly-excited states.}
	\label{fig:singlet}
\end{figure*}

The distribution of errors in the
excitation energy for the three methods assessed here is depicted in Fig.~\ref{fig:singlet},
and the corresponding statistical analysis is reported in Table \ref{tab:singlet}.
For the full set of singly-excited states, \Dhf and \Doo share a near-zero MSE (\SI{0.02}{\eV}),
and a comparable accuracy, with a small advantage for the latter (MAE of \SI{0.15}{\eV} against \SI{0.17}{\eV}).
\Doo produces fewer outliers than \Dhf, which is clear from their corresponding RMSEs, of \SI{0.27}{\eV} and \SI{0.35}{\eV}.
\Doo is also more consistent than \Dhf, having SDEs of \SI{0.26}{\eV} and \SI{0.35}{\eV}.
Orbital optimization enhances the description of singly-excited states, particularly for those poorly characterized by \Dhf, providing a more accurate overall representation.
Both methods compare unfavorably with \eom, however,
which has a considerably smaller MAE (\SI{0.08}{\eV}), RMSE and SDE, although it shows a somewhat positive MSE (\SI{0.05}{\eV}).
Compared to the closer performances of \Dhf, \Doo, and \eom for doublet-doublet transitions,
the larger discrepancies for singly-excited states are noteworthy.

%%% TABLE III %%%
\begin{table*}
\caption{Statistical results associated with the errors in excitation energies
(with respect to the reference values) for the singlet and triplet singly-excited states
computed with \eom, \Dhf, and \Doo.
Values in parenthesis are obtained by discarding 19 states having four dominant determinants with the exact same weight in the EOM-CCSD vector,
which are all $\pi \pi^*$ valence excitations involving degenerate orbitals. All values are in \si{\eV}.}
   \label{tab:singlet}
   \begin{ruledtabular}
       %\begin{tabular}{llrdddd}
       \begin{tabular}{llrrrrr}
            Method & Character &  \# states   &  \tabc{MAE}   &  \tabc{MSE}   & \tabc{RMSE}   &  \tabc{SDE} \\
            \hline
            \eom & All     &  231 (212) &   0.08 (0.09) &   0.05 ( 0.05) &   0.13 (0.13) &   0.12 (0.12) \\
                 & Singlet &  125 (116) &   0.10 (0.10) &   0.09 ( 0.09) &   0.13 (0.13) &   0.09 (0.10) \\
                 & Triplet &  106 ( 96) &   0.07 (0.06) &  -0.01 (-0.00) &   0.12 (0.13) &   0.12 (0.13) \\
                 & Valence &  152 (133) &   0.09 (0.09) &   0.04 ( 0.05) &   0.14 (0.14) &   0.13 (0.14) \\
                 & Rydberg &   79 ( 79) &   0.08 (0.08) &   0.05 ( 0.05) &   0.09 (0.09) &   0.08 (0.08) \\ \\

            \Dhf & All     &  231 (212) &   0.17 (0.13) &   0.02 (-0.01) &   0.35 (0.28) &   0.35 (0.28) \\
                 & Singlet &  125 (116) &   0.17 (0.11) &   0.03 (-0.02) &   0.38 (0.29) &   0.38 (0.29) \\
                 & Triplet &  106 ( 96) &   0.18 (0.15) &  -0.01 (-0.01) &   0.32 (0.28) &   0.32 (0.28) \\
                 & Valence &  152 (133) &   0.20 (0.14) &   0.06 ( 0.02) &   0.36 (0.24) &   0.35 (0.24) \\
                 & Rydberg &   79 ( 79) &   0.12 (0.12) &  -0.07 (-0.07) &   0.34 (0.34) &   0.33 (0.33) \\ \\

            \Doo & All     &  231 (212) &   0.15 (0.12) &   0.02 (-0.01) &   0.27 (0.20) &   0.26 (0.20) \\
                 & Singlet &  125 (116) &   0.15 (0.11) &   0.04 ( 0.00) &   0.29 (0.18) &   0.29 (0.18) \\
                 & Triplet &  106 ( 96) &   0.13 (0.12) &  -0.00 (-0.03) &   0.23 (0.22) &   0.23 (0.22) \\
                 & Valence &  152 (133) &   0.17 (0.13) &   0.04 (-0.02) &   0.31 (0.22) &   0.31 (0.22) \\
                 & Rydberg &   79 ( 79) &   0.09 (0.09) &  -0.01 (-0.01) &   0.15 (0.15) &   0.15 (0.15)
       \end{tabular}
   \end{ruledtabular}
\end{table*}

\Dhf and \Doo show a substantial number of absolute errors
exceeding \SI{0.2}{\eV}, in contrast to \eom. 
This significantly deteriorates the statistics of these methods, since
19 and 15 states, respectively, have errors above \SI{0.5}{\eV}.
For \eom, this happens for 2 states only.
Understanding the reasons behind the inaccuracy of \Dhf and \Doo for these specific
cases is important.
TD-CC is expected to perform well for states described by a single CSF; however, it is less effective for excited states that exhibit a pronounced multiconfigurational character, which would necessitate additional CSFs.
To evaluate its performance for the former type of states only, we excluded
19 states having four dominant determinants with the exact same weight in the EOM-CCSD vector, which are indicated in red in the \SupMat.
These transitions involve degenerate $\pi$/$\pi^*$ orbitals and would require at least two CSFs to be qualitatively described.
The associated statistics excluding results from this subset of singly-excited states are shown in parenthesis in Table \ref{tab:singlet}.
By discarding these multiconfigurational excited states, the statistics for both \Dhf and \Doo are systematically improved.
In particular, the MAE of \Dhf decreases from \SI{0.17}{\eV} to \SI{0.13}{\eV}, whereas that of \Doo decreases from \SI{0.15}{\eV} to \SI{0.12}{\eV}.
The accuracy of $\Delta$CCSD methods thus approaches that of \eom, which remains almost unaltered (with a slightly larger MAE of \SI{0.09}{\eV}) in this case.
Moreover, the comparable improvement for both \Dhf and \Doo suggests that orbital optimization is overall equally helpful for the single and multiconfigurational excited states.

Further examining the
\eom eigenvectors shows that \Dhf and \Doo usually deliver larger errors for states with
important contributions coming from several determinants, especially when these determinants
are doubly excited with respect to the two determinants considered
for the TD-CC calculation. From this last point, the magnitude of the errors becomes less surprising,
since we perform all single and double excitations on top of a two-determinant reference.
For instance, by further discarding the three excited states having the largest contribution from doubly-excited determinants with respect
to those used in TD-CC (based on the EOM-CCSD vector),
namely the ${1}^3A_1$ state of cyclopentadiene, the ${2}^1B_2$ state of cyclopropenone, and the ${1}^3A_g$ state of butadiene, highlighted in yellow in the \SupMat,
one obtains even smaller MAEs for \Dhf and \Doo, of \SI{0.12}{\eV} and \SI{0.11}{\eV}, whereas the MAE of \eom remains unchanged at \SI{0.09}{\eV}.
Clearly, the performance of $\Delta$CC methods, TD-CC in particular, approaches the one of \eom for states dominated by a single open-shell CSF but deteriorates as the multiconfigurational character of the excited states becomes more pronounced.

\begin{figure*}
    \includegraphics[width=0.32\textwidth]{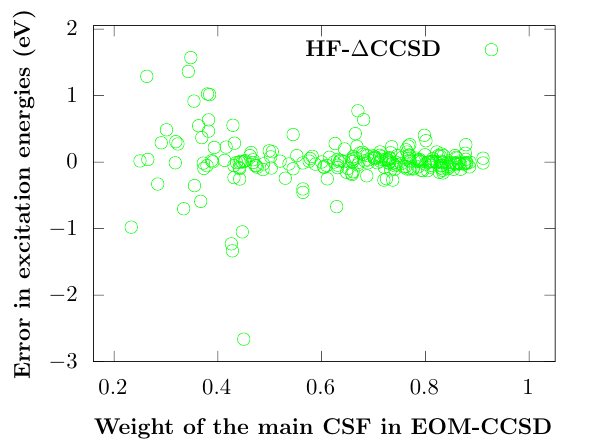}
    \includegraphics[width=0.32\textwidth]{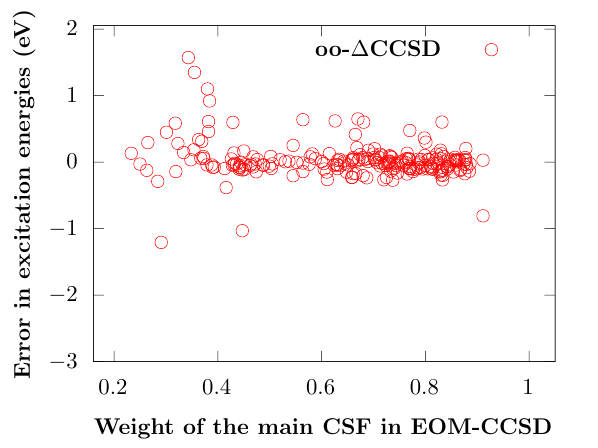}
    \includegraphics[width=0.32\textwidth]{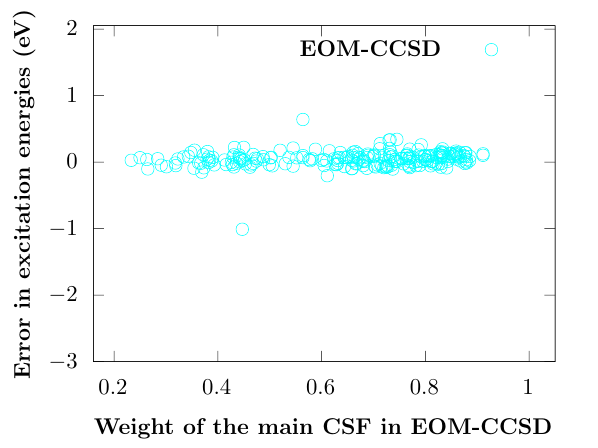}
    \caption{\alert{Scatter plots of the error in excitation energies (with respect to 
reference values) computed at the HF-$\Delta$CCSD (left), oo-$\Delta$CCSD (center),
and EOM-CCSD (right) levels, for open-shell singly excited states, as functions of
	the weight, in the EOM-CCSD vector, of the CSF employed as a reference in the TD-CCSD calculation.}
	}
    \label{fig:errors}
\end{figure*}

\begin{figure}
    \includegraphics[width=0.48\textwidth]{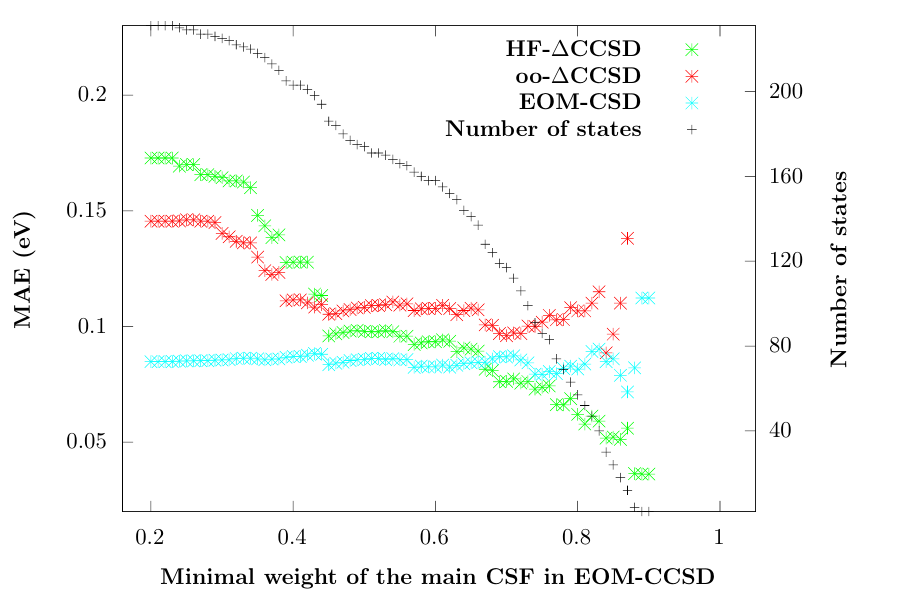}
    \caption{\alert{MAE of the excitation energy errors of \Dhf, \Doo, and \eom, computed on subsets of states for which the 
predominant CSF employed as a reference in the TD-CCSD calculation has a weight higher than a given value.}}
    \label{fig:mae}
\end{figure}

\alert{The latter point is illustrated in Fig.~\ref{fig:errors}, where we inspect
the evolution of the excitation energy errors as a function of the 
weight, in the EOM-CCSD vector, of the CSF employed as a reference in the TD-CCSD calculation. 
It becomes evident that the accuracy of the excitation energies deteriorates as the weight of 
the predominant CSF decreases. Similarly, the analysis of the same quantities at the \eom level
evidences that \eom manages to more accurately describe multi-CSF states. 
Complementarily, if we now look in Fig.~\ref{fig:mae} at the evolution
of the MAE of the excitation energies computed on the subset of states for which the 
predominant CSF has a weight higher than a given value, we find distinct trends for
TD-CCSD and EOM-CCSD. If we keep all the states, TD-CCSD has a large MAE and underperforms EOM-CCSD. If we retain only the states
with a weight above 0.45, HF-$\Delta$CCSD and EOM-CCSD exhibit very similar MAEs, while HF-$\Delta$CCSD becomes
more accurate than EOM-CCSD when the weight of the states exceeds 0.67.
However, caution is warranted for the final data points where 
the weight of the primary CSF exceeds 0.85, due to the limited number of states in this region.
Consequently, \Dhf outperforms \eom only for states adequately represented by a single CSF.}

\alert{In contrast, the performance of \Doo differs significantly. While it yields lower MAEs than 
\Dhf in the region of low weights, it exhibits minimal improvement as weights increase and never surpasses \eom. 
This discrepancy is likely attributable to the presence of states with substantial 
excitation energy errors despite a high weight assigned to the considered CSF, as depicted in Fig.~\ref{fig:errors}.
It may be related to difficulties in targeting the state of interest.}

\alert{Although the previous trends are likely valid for doublet-doublet transitions, 
we were unable to draw definitive conclusions due to the limited number of states available for analysis.}

Another important aspect concerns the accuracy of the $\Delta$CCSD methods with respect to the
nature of the excitation. As can be seen from Table \ref{tab:singlet},
\eom suffers from significant differences in its MAE and MSE when comparing singlet
and triplet states. This behavior may be attributed to variations in the $\%T_1$
values, as triplets generally have typically much higher $\%T_1$ than
singlets. \cite{Veril_2021} Thus, it is reasonable to expect a more favorable
treatment of triplets with \eom. On the other hand, \Dhf and \Doo share a similar MAE
and MSE, meaning that $\Delta$CCSD demonstrates greater consistency in the treatment
of singlet and triplet states.
When disregarding the more problematic multiconfigurational excited states,
the small gap between the MAE for singlets and triplets is preserved in the case of \Doo,
but actually increases for \Dhf, becoming comparable to that of \eom.

When looking at the statistics
for valence and Rydberg states, $\Delta$CCSD seems to be less consistent than \eom.
While it provides a better treatment of the Rydberg states, the description of
the valence ones is less satisfactory. The same trend is observed in \eom, though to a lesser degree. A closer examination of the statistics
reveals that the issue primarily arises from the $\pi\pi^*$ transitions,
which are less accurately described than the other valence excitations (the results are shown in the \SupMat).
However, the gap between valence and Rydberg transitions encountered for $\Delta$CCSD is considerably reduced
once the 19 multiconfigurational excited states described above are discarded (see Table \ref{tab:singlet}).
In fact,  \eom also exhibits a poorer description of $\pi\pi^*$ transitions, albeit by a smaller margin.

\Doo is also more accurate than $\Delta$CISD \cite{Kossoski_2023} for the singly-excited states,
with respective MAEs of \SI{0.10}{\eV} and \SI{0.16}{\eV}, respectively.
To have a fair comparison, these MAEs were computed by considering the subset of 203 excited states described by a single CSF reference in Ref.~\onlinecite{Kossoski_2023}.
In this case, the two sets of calculations only differ by the connectedness of CC, which implicitly accounts for higher-order excitations.
The \SI{0.06}{\eV} improvement equals the one found for the doublet-doublet transitions.

%%%%%%%%%%%%%%%%%%%%%%%%%%%%%%%%%%%%%%%%%%%%%%
\section{Concluding Remarks}
\label{sec:ccl}
%%%%%%%%%%%%%%%%%%%%%%%%%%%%%%%%%%%%%%%%%%%%%%

In this work, we studied the use of state-specific CCSD approaches to compute vertical excitation energies and compare them with the standard EOM-CCSD approach. Thanks to the present implementation of TD-CCSD and CCSD based on a non-Aufbau determinant, we conducted a substantial number of calculations on various types of excited states to assess the performance of these state-specific approaches.
Our investigation began with closed-shell doubly-excited states, where $\Delta$CCSD is known to work quite well (see, for example, Ref.~\onlinecite{Lee_2019}). Next, we extended our exploration to doublet-doublet transitions in radicals, for which both ground and excited states are generally well-described by a single (open-shell) Slater determinant. Finally, we inspected more traditional single excitation processes leading to both singlet and triplet excited states using TD-CCSD, which allows us to perform CCSD calculations on top of a single CSF.

In the context of doubly-excited states, $\Delta$CCSD produces results that are comparable to EOM-CCSDT and significantly more accurate than \eom. This is an expected outcome because, by construction, at the \eom level, the similarity-transformed Hamiltonian only contains the reference determinant in addition to its single and double excitations. This level of description is simply insufficient to accurately describe the genuine doubly-excited states considered here, which require the inclusion of triples and quadruples to be properly correlated.

The results for the other kinds of states yielded somewhat unexpected outcomes. Indeed, because \eom is biased toward the ground state, one might expect improved results by starting from a more suitable, state-specific reference. However, our findings did not indicate any substantial improvement with the state-specific approaches followed here. Additionally, employing optimized orbitals for the reference CSF in \Doo did not lead to a significant improvement in the accuracy when compared to \Dhf. This observation may be attributed to the fact that our optimized orbitals minimize the energy associated with the reference CSF but do not minimize the corresponding CCSD energy. Nonetheless, it is noteworthy that, for most of the states investigated here, $\Delta$CCSD did not significantly deteriorate the quality of the doublet states as compared to \eom, with a small MAE increase from \SI{0.07}{\eV} to \SI{0.10}{\eV} (for \Doo).

Finally, for singlet and triplet singly-excited states, $\Delta$CCSD produced
twice larger errors (MAEs of \SIrange{0.15}{0.17}{\eV}) than \eom (MAE of
\SI{0.08}{\eV}). Excited states with a strong multiconfigurational character
are less well described with $\Delta$CCSD approaches, which could be expected.
Ignoring \alert{the most pathological multi-reference} states from
the statistics reduces the MAEs by \SIrange{0.03}{0.04}{\eV}, which is however
not enough to match the systematically very good performance of \eom for the
states considered here. \alert{However, we demonstrate that for states characterized 
by a significant weight on a single CSF, $\Delta$CCSD yields more accurate results than \eom. 
Hence, it is important to note that $\Delta$CCSD is best suited for single-reference problems compared to \eom.}
We also found that state-specifically optimizing the
orbitals leads to a small but statistically significant improvement for this
class of excitation. On the bright side, $\Delta$CCSD enables a more consistent
treatment of singlet and triplet states than \eom.
Consequently, $\Delta$CCSD may offer an
attractive alternative for achieving a more consistent treatment of excited states
of different spin multiplicities, which remains to be explored. State-specifically
optimizing the ground- and excited-state orbitals at the CCSD level is another 
avenue one should explore but is left as a future work.

In comparison to standard multiconfigurational methods, such as CASPT2 or NEVPT2, $\Delta$CCSD demonstrates interesting performance. While the former methods typically yield excitation energy errors in the range of \SIrange{0.1}{0.2}{\eV}\cite{Schapiro_2013,Sarkar_2022,Boggio-Pasqua_2022,Battaglia_2023} and involve a computational complexity that grows factorially with the size of the active space, $\Delta$CCSD provides excitation energies with errors between \SI{0.10}{\eV} and \SI{0.15}{\eV} at a computational cost scaling polynomially with the system size. Thus, $\Delta$CCSD may emerge as an appealing alternative for accurately describing excitations that demand a large active space in multiconfigurational methods.

In addition to the classes of excited states considered in the present benchmark study, there remains the unexplored class of charge-transfer states, which could be of particular interest for future assessments of $\Delta$CCSD. \cite{Kozma_2020,Loos_2021c} In this case, orbital relaxation is known to be crucial, \cite{Schmerwitz_2023} and thus $\Delta$CCSD may be expected to outperform \eom. In addition, the present analysis mainly focused on small- and medium-sized molecules. Future investigations could extend this study to larger systems. Indeed, the quality of \eom deteriorates with increasing system size and becomes less accurate than EOM-CC2 for large systems. \cite{Loos_2021b} It remains to be seen whether this observation holds true within the state-specific formalism.

%%%%%%%%%%%%%%%%%%%%%%
\section*{Supporting Information}
\label{sec:supmat}
%%%%%%%%%%%%%%%%%%%%%%
See the {\SupMat} for the geometries of borole and oxalyl fluoride, further statistical measures, the TD-CCSD equations, the raw data associated with each figure and table, in addition to the energies at different levels of theory in the considered basis sets.

%%%%%%%%%%%%%%%%%%%%%%%%
\acknowledgements{
The authors thank Ajith Perera, Moneesha Ravi, P\'eter Szalay, and Rodney Bartlett for helpful discussions. This project has received funding from the European Research Council (ERC) under the European Union's Horizon 2020 research and innovation programme (Grant agreement No.~863481). This work used the HPC resources from CALMIP (Toulouse) under allocation 2023-18005 and from the CCIPL/GliCID mesocenter installed in Nantes.}
%%%%%%%%%%%%%%%%%%%%%%%%

%%%%%%%%%%%%%%%%%%%%%%%%%%%%%%%%
%\section*{Data availability statement}
%%%%%%%%%%%%%%%%%%%%%%%%%%%%%%%%
%The data that supports the findings of this study are available within the article and its supplementary material.

%%%%%%%%%%%%%%%%%%%%%%%%%%%%%%%%
\bibliography{state_specific_cc}
%%%%%%%%%%%%%%%%%%%%%%%%%%%%%%%%
\end{document}